\documentclass[10pt,twocolumn,twoside]{IEEEtran}
\newcommand{\comm}[1]{}
\def\xxxonly{\comm}
\def\xxxonly{ }
\def\noxxx{\comm}

\usepackage{color}

\usepackage[numbers]{natbib}\usepackage{graphicx} \usepackage{epsfig}\def\citet{\cite}

\usepackage{amssymb}
\usepackage{setspace}
\usepackage{ae} 
\usepackage{eucal} 

\newtheorem{theorem}{Theorem}
\newtheorem{lemma}{Lemma}

\newtheorem{proposition}{Proposition}

\newtheorem{definition}{Definition}
\newtheorem{remark}{Remark}

\newtheorem{example}{Example}[section]

\def\e{\varepsilon}

\def\defi{\stackrel{{\scriptscriptstyle \Delta}}{=}}

\def\a{\alpha}

\def\o{\omega}

\def\Y{{\cal Y}}

\def\w{\widehat}
\def\Ind{{\mathbb{I}}}

\def\esssup{\mathop{\rm ess\, sup}}

\def\Z{{\cal Z}}

\def\H{{\cal H}}

\def\h{h}
\def\L{L}

\def\C{{\bf C}}

\def\ww{\widetilde}
\def\X{{\cal X}}

\def\oo{\bar}

\def\U{{\cal U}}

\def\L{{\cal L}}

\def\h{h}

\newcommand{\be}{\begin{equation}}
\newcommand{\ee}{\end{equation}}
\newcommand{\bd}{\begin{displaymath}}
\newcommand{\ed}{\end{displaymath}}
\newcommand{\ba}{\begin{array}{ll}}
\newcommand{\ea}{\end{array}}
\newcommand{\baa}{\begin{eqnarray}}
\newcommand{\eaa}{\end{eqnarray}}
\newcommand{\baaa}{\begin{eqnarray*}}
\newcommand{\eaaa}{\end{eqnarray*}}   \font\sm=cmr10

\def\k{\kappa}

\def\H{{\cal H}}

\def\SS{\bar\Sigma}

\def\e{\varepsilon}

\def\defi{\stackrel{{\scriptscriptstyle \Delta}}{=}}

\def\a{\alpha}

\def\o{\omega}

\def\Y{{\cal Y}}

\def\w{\widehat}
\def\Ind{{\mathbb{I}}}

\def\Z{{\cal Z}}
\def\ZZ{{\bf Z}}

\def\H{{\cal H}}

\def\h{h}

\def\C{{\bf C}}

\def\T{{\mathbb{T}}}
\def\TT{{\cal T}}
\def\ZZ{{\mathbb{Z}}}
\def\a{\alpha}

\def\ew{\left(e^{i\o}\right)}
\def\L{{\cal L}}

\def\NN{{\scriptscriptstyle N}}
\def\BL{{\scriptscriptstyle BL}} 
\def\WW{\varkappa}
\def\WW{{w_n}}
\def\wh{\ww h_n}
\def\wH{\ww H_n}
\def\wHH{\H_{\TT}}\def\EE{{\mathbb{E}}}
\title{On recovery of signals  with single point spectrum degeneracy }
\author{
Nikolai Dokuchaev }
\begin{document}
\def\break{}%
\def\brea{}
\def\breakk{}
\def\brea{\nonumber\\ }\def\breakk{\nonumber\\&&} 
\maketitle
\let\thefootnote\relax\footnote{\xxxonly{Submitted:  September 24, 2018.}
\par
The author is with  School of Electrical Engineering, Computing and Mathematical Sciences, Curtin
University,   GPO Box U1987, Perth, 6845 Western Australia\noxxx{  and also with National Research University ITMO, 197101 Russia}. }

\begin{abstract}
The paper study recovery problem for discrete time signals
 with a finite number of missing values.  The paper establishes
 recoverability of these missing values for signals with Z-transform
  vanishing with a certain rate at a single  point.
  The transfer functions for the corresponding recovering kernels  are presented  explicitly.
  Some robustness of the recovery with respect to data truncation or noise contamination is established.
  \par
{\bf Key words}: data recovery, spectrum degeneracy, Z-transform, transform functions, robustness.
\end{abstract}
\section{Introduction}
\noxxx{
A core problem of the mathematical theory of signal processing is the problem
of recovery of missing  data. For continuous data, the recoverability is associated with smoothness or analytical properties of
the processes. For
discrete time processes,  it is less obvious how to interpret analyticity; so far, these problems
were studied in a stochastic setting,
where an observed process is deemed to be  representative of an ensemble of paths with
the probability distribution that is either known or can be estimated from repeating experiments.
A classical result for stochastic stationary  Gaussian processes
with the  spectral density $\phi$ is that
a missing  single value is recoverable  with zero
error if and only if \baa
\int_{-\pi}^\pi \phi(\o)^{-1} d\o=-\infty.\label{Km} \eaa
(Kolmogorov \cite{K},
Theorem 24). Stochastic stationary Gaussian processes without this property are called {\em minimal } \cite{K}.
Criterion (\ref{Km}) was extended on stable processes \citet{Peller}  and vector Gaussian processes  \citet{Pou,Pou2}.
Clearly, (\ref{Km}) holds for all ``band-limited" processes meaning that the spectral density is vanishing on an arc of the
unit circle $\T=\{z\in\C:\ |z|=1\}$.

It is known that signals with certain restrictions on the spectrum or sparsity  feature recoverability of missing data in the pathwise setting without probabilistic assumptions.
This setting targets situations where we deal with a sole sequence that is deemed to be unique
  and such that one cannot rely on statistics  collected from observations of other similar samples.  An estimate of the missing  value
   has to be done based  on the intrinsic properties of  this sole  sequence and the observed values.
 For example,  a subsequence  sampled at sparse enough periodic points   can be removed from observations of an  oversampling sequence \cite{F95}.\index{(Ferreira (1995)).}  In the compressive sensing    setting,
 reducing of the sampling rate for finite sequences has been achieved using sparsity of signals \cite{Donoho,CJR}. The connection of
 bandlimiteness and  recoverability  from  samples was established  for
 the fractional Fourier transform
 \cite{F3}. There is also a so-called  Papoulis approach \cite{Pa} allowing to reduce the  sampling rate with additional measurements at sampling points; this approach was extended on multidimensional processes \cite{Ch1}.

There is also an approach based on the so-called Landau's phenomenon \cite{La,La2};, it was shown in  \cite{La} that there is an uniqueness  set of sampling points representing small deviations of integers for classes of functions with an arbitrarily large measure of the spectrum range.
This result was extended on functions with unbounded spectrum range and on sampling points
 allowing a convenient  explicit representation \cite{OU08}.The paper suggests  a criterion of   recoverability of sequences (discrete time processes) with spectrum degeneracy at a single point.
  The result is based  on the approach developed for pathwise predicting  \citet{D12a,D12b}, where  some predictors were derived to
  establish  arbitrarily precise predictability.  In the present paper, arbitrarily precise recoverability is  established for certain classes of  square-summable sequences (processes) with Z-transform
  vanishing at a single point  (Theorem \ref{ThM});  the sequences are not necessarily summable.
  The required decay rate is mild.
   The corresponding recovering kernels are obtained and represented
  explicitly via  their transfer functions. Some robustness with respect to noise contamination is established for the suggested recovering algorithm.
Related results were obtained in \cite{D16,D17}. In \cite{D16}, the setting for recovery of $M$ missing values covers processes with
Z-transforms vanishing  at $M$ points located periodically on $\T$.
In \cite{D17}, the result covers only the cases of sequences from $\ell_1$ with one-point spectrum degeneracy  or sequences
from $\ell_2$ that are
band-limited, i.e. with Z-transform vanishing on an  arc of the unit circle $\T$.
The corresponding spectrum degeneracy would require that $M$ first derivatives of $X\ew$ vanish at $\o=\pi$; here $X$ is the Z-transform of an underlying process. The approach from \citet{D17} is not applicable to processes  $\ell_2\setminus \ell_1$
with spectrum degeneracy at a single point, since the recovery kernels used therein were non-vanishing sequences from $\ell_\infty\setminus \ell_2$.
The approach of the present paper is quite different form the one from \cite{D16,D17} and does not involve restrictions on the derivatives of Z-transforms or spectrum degeneracy at multiple points.
}
\xxxonly{ The paper presents a modification of the approach developed in \cite{D16}
for recovery of a finite set of missing values for discrete time signals. This important problem was widely studied; see the literature review in \cite{D16}.

The result and the approach of the present
paper are different from the result  \cite{D16}. In particular,  the conditions of recoverability and the 
recovering kernels obtained in
   the present paper are different from the ones in \cite{D16}.
   Overall, the  result of the present paper  has some advantages as well as some disadvantages
comparing with \cite{D16}. In particular, it has the following advantages.
\begin{enumerate}
 \item In \cite{D16},  sufficient conditions of recoverability of $M$ missing values require spectrum degeneracy of the underlying processes at $M$ isolated points located periodically on the unit circle in $\{e^{i\o}:\ \o\in (-\pi,\pi]\}$. In the present paper, the spectrum degeneracy is required at  a single point $e^{i\pi}=-1$ only.
   \item In \cite{D16},  the  values can be missed  for consequent times only.
   In the present paper, the times for missing  values can be scattered arbitrarily.
   Moreover, the method developed in the present paper is more efficient for the case large distance between the times for missing values.
\end{enumerate}
However, the result of the present result has the following  disadvantages
comparing with \cite{D16}.
\begin{enumerate}
 \item For the case of $M>1$, the rate of spectrum degeneracy required in \cite{D16}
   at  each of the points of degeneracy is lower comparing with the rate of degeneracy at $|\o|=\pi$  required in the present paper.
  \item The rate of the spectrum degeneracy required in \cite{D16}
   at  each of the points of degeneracy does not depend on the number $M$ of missing points.
  The rate of the spectrum degeneracy at $e^{i\pi}=-1$  required in the present paper is increasing with $M$.
   \item The rate of the spectrum degeneracy required in \cite{D16}    is described explicitly. In the present
   paper, the rate of the  spectrum degeneracy has to be calculated via a numerical procedure.
       \end{enumerate}
\section{Definitions and background}
Let $\T\defi\{z\in\C:\ |z|=1\}$, and let $\ZZ$ be the set of all
integers.
\par
We denote by $\ell_r$ the set of all sequences
$x=\{x(t)\}\subset\C$, $t=0,\pm 1,\pm 2,...$, such that
$\|x\|_{\ell_r}=\left(\sum_{t=-\infty}^{\infty}|x(t)|^r\right)^{1/r}<+\infty$
for $r\in[1,\infty)$ or  $\|x\|_{\ell_\infty}=\sup_t|x(t)|<+\infty$
for $r=+\infty$.
\par
For  $x\in \ell_1$ or $x\in \ell_2$, we denote by $X=\Z x$ the
Z-transform  \baaa X(z)=\sum_{t=-\infty}^{\infty}x(t)z^{-t},\quad
z\in\C. \eaaa Respectively, the inverse $x=\Z^{-1}X$ is defined as
\baaa x(t)=\frac{1}{2\pi}\int_{-\pi}^\pi X\left(e^{i\o}\right)
e^{i\o t}d\o, \quad t=0,\pm 1,\pm 2,....\eaaa

We have that  $x\in \ell_2$ if and only if  $\|X\ew\|_{L_2(-\pi,\pi)}<+\infty$.   In addition, $\|x\|_{\ell_\infty}\le \|X\ew\|_{L_1(-\pi,\pi)}$.

We use the sign $\circ$ for convolution in $\ell_2$.

For a finite set $S$, we denote by $|S|$ the number of its elements.

\subsection*{The setting for the recovery problem}
Let a finite set $\TT\subset\ZZ$ be given.

Let  $\wHH$ be the set  of all $h\in\ell_2$
such that, for all $x\in\ell_2$,
\baa
(h\circ x)(t)=\sum_{s\in\ZZ:\ s\notin \TT}h(t-s)x(s),\quad t\in\TT.
\label{wHH}
\eaa

We are interested  in the problem of recovery  values
$\{x(t)\}_{t\in\TT}$  from  observations $\{x(s)\}_{s:\ s\notin\TT}$ for $x\in\ell_2$.
More precisely,  we consider  calculation of estimates $\{\w x(t)\}_{t\in\TT}$
obtained  as $\w x=h\circ x$  for some appropriate kernels $h\in\wHH$.
\begin{definition}\label{def} Let $\Y\subset \ell_2$ be a class of sequences.
\begin{itemize}
\item[(i)] We say that this class is  recoverable if
there exists a sequence $\{h_n(\cdot)\}_{n=1}^{+\infty}\subset
\wHH$ and \baaa \sup_{t\in\ZZ}|x(t)-\w x_n(t)|\to 0\quad
\hbox{as}\quad m\to+\infty\quad\forall x\in\Y, \eaaa where $\w x_n=h_n\circ x$.
\item[(ii)]
 We say that the class $\Y$ is  uniformly recoverable  if, for any $\e>0$, there exists $h(\cdot)\in \wHH$ such that \baaa \sup_{t\in\ZZ} |x(t)- \w x(t)|\le \e\quad
\forall x\in\Y, \label{predu}\eaaa where $\w x=h\circ x$.
\end{itemize}
\end{definition}

Since $h\circ x=\Z (H X)$, where $H=\Z h$ and $X=\Z x$, then it follows that desired recovery operators should have the following properties.
 \begin{itemize}
 \item[(a)] $h\in \H$;
 \item[(b)] $H\ew\cdot X\ew\approx X\ew$.
 \end{itemize}
We will show below that  this can be satisfied for appropriate choice of $h$ and for some wide enough classes of processes.
\section{The main results}
We will establish recoverability for sequences  with Z-transforms vanishing at some isolated  points of $\T$ with certain rate.
\subsection{Recoverability for some classes of processes}
For $n=1,2,...$, let $D_n\defi (\pi-\pi/n,\pi)$ and $\ww D_n\defi (-\pi,-\pi+\pi/n)\cup (\pi-\pi/n,\pi)$.

For $t\in\TT$, let $S_t=\{k\in\ZZ:\  k=t-s,\ s\in\TT\}$. Let $S_{\TT}=\cup_{t\in\TT}S_t$.
It can be noted that $0\in S_{\TT}$ for any choice of $\TT$; in addition,  if $s\in S_\TT$ then $-s\in S_\TT$.
\begin{proposition}\label{propM} Let $h\in\ell_2$ and $r\in\ZZ$. Then   $h\in\wHH$ if and only if $h(t)=0$ for all $t\in S_{\TT}$.
\end{proposition}
\begin{example}\label{exS} \begin{enumerate}
\item
For  $\TT=\{t=1,3,4\}$, we have that $S_{\TT}=\{-3,-2,-1,0,1,2,3\}$.
\item
Let an integer $m\ge 0$ be given.
For  $\TT=\{t\in\ZZ:\ |t|\le m\}$, we have that $S_{\TT}=\{s\in\ZZ:\ |s|\le 2m\}$.
\item
Let an integer $m\ge 0$ be given.
For  $\TT=\{t=0,1,...,m\}$, we have that $S_{\TT}=\{s\in\ZZ: \ |s|\le m\}$.
\end{enumerate}
In this example, the ratio  $|S_{\TT}|/|\TT|$ is the same for the cases (ii) and (iii).
This ratio is larger for the case where  the set $\TT$ is non-periodic; in particular,
this ratio is larger for the case (i) then for the case (ii) with $m=1$ or for the case  (ii) with $m=2$
(however, the number of missing  values is the same for all three case).
 \end{example}

\def\SS{P}
Let $\SS_{\TT}\defi\{s\in S_{\TT}:\ s> 0\}$.

For an integer $n>1$, let $\SS_{n,\TT}\defi\{s\in \SS_{\TT}:\ s/n\in\ZZ\}$
and let  $\oo \SS_{n,\TT}\defi \SS_{\TT}\setminus \SS_{n,\TT}$.

Let elements of $\SS_{\TT}$ be counted as $\{t_1,....,t_{p},t_{p+1},...,t_q\}$,
where
$p=|\SS_{n,\TT}|$ and  $q=|\oo \SS_{n,\TT}|+p$, such that
 $\SS_{n,\TT}=\{t_1,....,t_{p}\}$ and  $\oo \SS_{n,\TT}=\{t_{p+1},....,t_{q}\}$.

The case where $\oo \SS_{n,\TT}=\emptyset $ is not excluded; in this case, $q=p$.

Let   $e_k(\o)\defi \cos(t_k\o)$, $k=1,...,q$, and let $e_0(\o)\equiv 1$.

We consider $L_2(D_n)$
as  a real  Hilbert space with the standard
 norm $\|\cdot \|_{L_2(D_n)}$ and the inner product $(\cdot,\cdot)_{L_2(D_n)}$.

Let $\L(f_1,...,f_q)$ denote a linear span of a set of functions $\{f_k\}\subset L_2(D_n)$.

For  $n>0$,  let a set $\{v_n^{(k)}\}_{k=1}^{q}\subset \L(e_1,...,e_q)$
be constructed
using the Gram-Schmidt orthogonalisation procedure such that  \baaa
&&\|v_k\|_{L_2(D_n)}=1,\quad k=1,...,q,\qquad\breakk
(v_k,v_n^{(l)})_{L_2(D_n)}=0,\quad  k,l=1,...,q,\quad k\neq l.
\eaaa
We assume that this procedure is run consequently according to numbering for $e_k$; this ensures
that $v_k\in \L(e_1,...,e_p)$ for $k=1,...,p$.

Let
\baaa
&&
\xi_n\defi e_0-\sum_{k=p+1}^{q}(e_0,v_k)_{L_2(D_n)}v_k,\\
&&\WW\defi (\pi-\pi/n) \frac{\xi_n}{(\xi_n,e_0)_{L_2(D_n)} }.
\eaaa
\begin{proposition}\label{propxi}  $0<(\xi_n,e_0)_{L_2(D_n)}=\|\xi_n\|_{L_2(D_n)}^2\le n^{-1}$ and
$\|\WW\|_{L_2(D_n)}\ge (\pi-\pi/n)\sqrt{n}$.
\end{proposition}

Let $\X_\TT$ be the class of all sequences $x\in\ell_2$ such that, for  $X=\Z x$,
\baa \int_{\ww D_n}|\WW(\o) X\ew| d\o\to 0\quad \hbox{as}\quad n\to +\infty.
\label{hfin}
\eaa
\par
By Proposition \ref{propxi}, requirement (\ref{hfin})  implies that $X\ew \to 0$ as $|\o|\to\pi$ for $x\in\X_\TT$.
This holds for ``degenerate'' processes, with $X\ew$
vanishing as $|\o|=\pi$ with certain rate of decay. We call this single point $\WW$-degeneracy.
\begin{example}
Let $\X^\BL$ be the set of all $x\in\ell_2$ such that  there exists
$n>0$ such that $X\ew|_{\o\in \ww D_n}=0$ (in particular, these processes are band-limited).
Then $\X^\BL\subset\X_\TT$ for any $\TT$.
\end{example}
\begin{example}\label{ex/n}
Assume that $n$ and $\TT$ are such that  $t/n\in \ZZ$ for any $t\in \SS_{n,\TT}$. Then
 $\xi_n=e_0$ and $w_n(\o)\equiv (\pi-\pi/n)n$.
\end{example}

\begin{definition}\label{defU}
 Let
 $\U\subset \X_\TT$   be a class of sequences. We say that this class features
  single point $\WW$-degeneracy uniformly over $\U$ if, for any $x\in \U$,
 (\ref{hfin})  holds uniformly over $x\in \U$  for $X=\Z x$.\end{definition}
 \begin{theorem}\label{ThM} The following holds.
  \begin{itemize}
\item[(i)]
The class $\X_\TT$ is  recoverable in the sense of Definition
\ref{def}(i).
\item[(ii)] Any class $\U\subset\X_\TT$ featuring uniform degeneracy in the sense of Definition
\ref{defU}  is  recoverable uniformly on $\U$ in the sense of Definition
\ref{def}(ii).
\end{itemize}
\end{theorem}

\subsection{The recovering kernels}
We assume that $\WW(\o)$ is extended on $\ww D_n$ from $D_n$  such that
$\WW(\o)=\WW(-\o)$.
\begin{lemma}\label{lemmawH}
For $n>1$, consider  kernels   $\wh(\cdot)=\Z^{-1}\wH$,
where \baaa  &\wH\ew = 1,\quad &|\o| < \pi-\pi/n,
\\  &\wH\ew = -\WW(\o),\quad &|\o|\in [\pi-\pi/n,\pi).
\eaaa
Then these $\wh$ are  real valued processes such that
$\wh(t)=\wh(-t)$ for all $t\in \ZZ$, and
\baa
&& \wh(0)=0,\quad\nonumber
\\ &&\wh(t)=\frac{1}{\pi t}\sin\left(\pi t -\frac{t\pi}{n}\right),\quad t\in S_{\TT}\setminus\{0\}.
\label{wh}\eaa
\end{lemma}

\begin{lemma}\label{lemmaH}
For $n>1$, consider  kernels constructed as
\baaa
h_n(t)=\wh(t)\Ind_{t\notin S_{\TT}}.
\eaaa
Then $h_n\in\H$ and  $\|\wh-h_n\|_{\ell_p}\to 0$ as $n\to +\infty$ for $p\in[1, +\infty]$.
\end{lemma}

Let $H_n\defi \Z h_n$.

Since $\wH\ew= 1$  and $H_{n}\ew \approx 1$  on a large part of $(-\pi,\pi)$ for large $n\to +\infty$, the kernels introduced in Lemma \ref{lemmaH}  are potential candidates for the role of
recovering kernels presented in Definition \ref{def} and required for recoverability claimed
in Theorem \ref{ThM}.
The following theorem shows that these
kernels ensure required recoverability.
\begin{theorem}\label{Th}
The  kernels  $h_{n}$ introduced in Lemma \ref{lemmaH}
 ensure recovering required in Definition \ref{def} (i)-(ii)  as $n\to +\infty$.
\comm{This means that
 \baaa \sup_{t\in\ZZ}|x(t)-\w x(t)|\to 0\quad \hbox{as}\quad
n\to+\infty\quad\forall x\in\X_\TT. \eaaa Moreover, for any
$\e>0$, there exists $n>0$ such that \baa \sup_{t\in\ZZ}|x(t)- \w
x(t)|\le \e\quad \forall x\in\U. \label{pred}\eaa
Here  $\w x(t)\defi \sum_{s\in \ZZ}h_{n}(s)x(t-s).$}
\end{theorem}
\begin{example} Under the assumption of Example (\ref{ex/n}),
  $\|H_{n}\ew\|_{L_1(-\pi,\pi)}\le 4\pi$ and $\sup_n\|h_{n}\|_{\ell_2}\le 2$.
\end{example}

\begin{remark}
\label{wwlarge}
It can be  seen that, for a fixed $n>0$,  large $|S_\TT^+|$  (i.e. large $|\TT|$) leads to small $\|\xi_n\|_{L_2(D_n)}$ and large $\|w_n\|_{L_2(D_n)}$.
On the other hand, large
$t_1=\min_{s,t\in\TT:\ s\neq t}|s-t|$ leads to smaller $\|\WW\|_{L_2(D_n)}$ (i.e. closer  to $(\pi-\pi/n)\sqrt{n}$).
\end{remark}
\begin{remark}
Theorem \ref{ThM} implies  that, for a given $\TT$,  the recovery error can be made arbitrarily small  via increasing $n$.
On the other hand,  Proposition \ref{propxi} implies that $\|H_{n}\ew\|_{L_2(-\pi,\pi)}\ge \|w_{n}\|_{L_2(D_n)} \to +\infty$ and hence $\|h_{n}\|_{\ell_2}\to +\infty$ as $n\to +\infty$. This means  that the values  $|h_n(t)|$ are  decaying   as $t\to +\infty$ slower for large $n$ required for lesser recovery error. Therefore, more precise recovery would
be more impacted by data truncation and would require more observations, especially  for heavy tail inputs.
\end{remark}
\begin{remark}\label{corr1} For $h\in\H$, the   operators $\w x=h\circ x$ are such that
for any $t\in\ZZ$, the estimates
$\{\w x(t+r)\}_{r\in\TT}$  are calculated  from  the observations $\{x(t+r)\}_{r\notin \TT}$.
More precisely, for any $x\in\ell_2$ and any $h\in\wHH$,
\baaa
(h\circ x)(t+r)=\sum_{s\in\ZZ:\ s\notin r+\TT}h(t+r-s)x(s),\quad t\in\TT.
\label{hhr}
\eaaa
Here $r+\TT=\{r+t,\ t\in\TT\}$. Therefore, the recovery kernels $h_n$ also solve the recovery problem in related time-invariant setting.
\end{remark}

\section{Proofs}\label{secProof}
\par

{\em Proof of Proposition \ref{propM}}  follows immediately from the fact that the equality
\baaa
\sum_{s\in\ZZ}h(t-s)x(s)=\sum_{s\in\ZZ\setminus\TT}h(t-s)x(s).
\eaaa
holds for all $t\in\TT$ and $x\in\ell_2$ if and only if  $h(t-s)=0$ for all $t,s\in\TT$.
$\Box$

{\em Proof of Proposition \ref{propxi}}. The function
$v_\L\defi \sum_{k=1}^{d}(e_0,v_k)_{L_2(D_n)}v_k$
represents the projection of $e_0$ on $\L_n$.  Since  $e_0\notin \L_n$, it follows that $\xi_n=e-v_\L\neq 0$, i.e.
 $\|\xi_n\|_{L_2(D_n)}\neq 0$. Hence \baaa
 (\xi_n,e_0)_{L_2(D_n)}=(\xi_n,\xi_n)_{L_2(D_n)}+(\xi_n,v_\L)_{L_2(D_n)}\brea=\|\xi_n\|_{L_2(D_n)}
 \eaaa
and \baaa
\|\xi_n\|_{L_2(D_n)}\le \|e_0\|_{L_2(D_n)}=n^{-1/2}.
\eaaa
This implies  Proposition \ref{propxi}. $\Box$

{\it Proof of Lemma \ref{lemmawH}}.
Since $\wH\left(e^{-i\o}\right)=
\overline{\wH\left(e^{i\o}\right)}$, we have that $\wh$ and $h_n$ are real valued.

Since all values $\wH\left(e^{-i\o}\right)$ are real, we have that $h_{n}(t)=h_{n}(-t)$ for all $t\in \ZZ$.

For $k=1,..,p$, we have that $t_k/n\in\ZZ$ and hence
\baaa (e_0,e_k)_{L_2(D_n)}=\int_{\pi-\pi/n}^\pi \cos( t_k\o)d\o
\brea=-\frac{1}{t_k}\sin(t_k(\pi-\pi/n))=0.\eaaa
Hence
\baaa
(e_0,v_k)_{L_2(D_n)}=0,\quad k=1,...,p
\eaaa
and
\baaa
(\xi_n,v_k)_{L_2(D_n)}=0,\quad k=1,...,p.
\eaaa
\par By the choice of the $\xi_n$, it follows that
\baaa
(\xi_n,v_k)_{L_2(D_n)}=0,\quad k=p+1,...,q.
\eaaa
Hence
\baaa
(\xi_n,e_k)_{L_2(D_n)}=0\quad k=p+1,...,q.
\eaaa
Finally, we obtain that
\baaa
(\xi_n,e_k)_{L_2(D_n)}=0\quad k=1,...,q.
\eaaa
By the definition of $\WW$ and $e_k$,  it gives that
\baa
\int_{D_n}\WW(\o)\cos(t_k\o) d\o=0,\quad k=1,...,q.
\label{e=0}
\eaa
\par
Furthermore, we have that
\baaa
\int_{D_n}\WW(\o)d\o=(\WW,e_0)_{L_2(D_n)}=\pi-\pi/n
\eaaa
and
\baaa
\int_{\ww D_n}\WW(\o) d\o=2(\pi-\pi/n),\quad\brea \int_{\ww D_n}\wH\ew d\o=-2(\pi-\pi/n).
\eaaa

 On the other hand,
 \baaa
\int_{(-\pi,\pi]\setminus\ww D_n}\wH\ew d\o=\int_{-\pi+\pi/n}^{\pi-\pi/n}\wH\ew d\o\brea=2(\pi-\pi/n).
\eaaa
Hence
\baaa \wh(0)=\frac{1}{2\pi}\int_{-\pi}^\pi
\wH\ew d\o=0.\eaaa
By (\ref{e=0}), we have for $t\in \SS_\TT$ that
\baaa
&&\wh(t)=\frac{1}{2\pi}\int_{-\pi}^{\pi}\wH\ew e^{i\o t}d\o\breakk=\frac{1}{\pi}\int_{0}^{\pi}\wH\ew \cos(\o t)d\o
\\&&=\frac{1}{\pi}\int_{0}^{\pi-\pi/n}\wH\ew \cos(\o t) d\o\breakk-\frac{1}{\pi}\int_{D_n} \WW(t)\cos(\o t) d\o
\\&&=\frac{1}{\pi}\int_{0}^{\pi-\pi/n}\cos(\o t) d\o=\frac{1}{\pi t}\sin(t\pi-t\pi/n).\eaaa
Since $\wh(t)=\wh(-t)$, this gives (\ref{wh}) and completes the proof of  Lemma \ref{lemmaH}. $\Box$

\vspace{0.5cm}
\par
{\it Proof of Lemma \ref{lemmaH}}. It suffices to observe that $\|h_n-\wh\|_{\ell_p}=\|y\|_{\ell_p}$,
where $y(t)\defi \frac{1}{\pi t}\sin(t\pi-t\pi/n)\Ind_{t\in S_{\TT}\setminus \{0\}}\to 0$ for all $t$ as $n\to +\infty$. $\Box$
\par
{\it Proof of Theorem \ref{ThM}}. Let $n\to +\infty$,  and let $\wH$  be as defined in Lemma \ref{lemmawH}. Let
$x\in \X_\TT$, $X\defi \Z x$,  $\wh=\Z^{-1}\wH$, and
\baaa
\ww x(t)\defi \sum^t_{s=-\infty}\wh (t-s)x(s). \eaaa
By
the definitions, it follows that
 $\ww X\left(e^{i\o}\right)\defi \wH\left(e^{i\o}\right)X\left(e^{i\o}\right)=(\Z \w x )\ew$.
\par
\par
We have that \baaa
&&\|\ww X\ew-X\ew\|_{L_1 (-\pi,\pi)}\breakk=\int_{\ww D_n}|\ww X\ew-X\ew| d\o\\
&&=\int_{\ww D_n} |(\wH\ew-1) X\ew|  d\o
 \le \zeta_n+\psi_n,
\eaaa
where
\baaa
\zeta_n=&&\|\WW(\o) X\ew
\|_{L_1(\ww D_n)},\qquad
\breakk \psi_n=\left\|X\ew \right\|_{L_2(\ww D_n)}.
\eaaa
By the assumptions on  $\X_\TT$,
\baaa
\zeta_n \to 0,\quad \psi_n \to 0\quad \hbox{as}\quad n\to +\infty.
\eaaa    Hence \baa
\|\ww x_n-x\|_{\ell_\infty}\to
0\quad\hbox{as}\quad n\to +\infty \quad \forall\in\X_\TT.\label{L1}
\eaa
Further,   let  $\h_n=\Z^{-1}\wH$   be  defined in Lemma \ref{lemmaH}, and let $\w x_n=h_n\circ x$.
 Let $H_n=\Z h_n$. By  Lemma \ref{lemmaH}, we have that
 \baaa
 \|h_n-\w h\|_{\ell_1} \to 0\quad \hbox{as}\quad n\to +\infty.
 \eaaa
 Hence
  \baa
 \esssup_{\o\in (-\pi,\pi]}|H_n\ew-\wH\ew|\to 0\quad \hbox{as}\quad n\to +\infty.
 \label{LH}\eaa
Let $x\in \X_\TT$ and  $X\defi \Z x$. We have that
\baaa
 \|x-\w x_n\|_{\ell_\infty}\le \|x-\ww x_n\|_{\ell_\infty}+ \|\w x-\ww x_n\|_{\ell_\infty}.
 \eaaa
We have that \baaa
&&\|\w x-\ww x_n\|_{\ell_\infty}\breakk=\sup_{t\in\ZZ}
\frac{1}{2\pi}\left|\int_{-\pi}^\pi \wH\ew -H_n\ew e^{i\o t} X\ew d\o\right|
\\ &&\le \frac{1}{2\pi} \esssup_{\o\in (-\pi,\pi]}|H_n\ew-\wH\ew \int_{-\pi}^\pi \left|X\ew\right| d\o.
 \eaaa
By (\ref{LH}), $\|\w x-\ww x_n\|_{\ell_\infty}\to 0$ as $n\to +\infty$. Together with (\ref{L1}), this implies that
$\|x-\w x_n\|_{\ell_\infty}\to 0$ as $n\to +\infty$.
 This
completes the proof of statement (i).

Let us prove statement (ii). By the assumptions on $\U$, it follows from the the proof above that    \baaa \|\w X\ew-X\ew\|_{L_1(-\pi,\pi)}\to 0\quad \hbox{as}\quad n\to +\infty
\eaaa uniformly over
$x\in\U$. In particular, for any $\e>0$, one can select $n$ such that
$\|\w x-x\|_{\ell_\infty}\le \e$.  This
completes the proof of statement (ii). It follows from the proofs above  that the
recovering kernels $\wh(\cdot)=\Z^{-1}\wH$ are such as required.
 This completes the proof of Theorem \ref{ThM}. $\Box$
 \def\NN{\eta}
 \def\nuu{\sigma}
\section{On robustness with respect to noise contamination}\label{secRob}
Let us discuss the impact of the presence
of the noise contaminating recoverable sequences.
Assume that the kernels $\wh$  described in Theorem \ref{ThM}  and
designed for recoverable sequences are
applied to a sequence with a noise
contamination.Let $\U\subset \X_\TT$  be a set such as described in Definition \ref{defU}.  Let us consider an input sequence $x\in\ell_2$
such that $x=x_0+\eta$, where $x_0\in\U$,
and where $\eta\in\ell_2\setminus \X_\TT$ represents a
noise.
 Let $X=\Z x$, $X_0=\Z x_0$, and $N=\Z \eta$. We
assume that
 $\|N\ew\|_{L_1(-\pi,\pi)}=\nuu$; the parameter $\nuu\ge 0$ represents the
intensity of the noise.
\par
In the proof of Theorem \ref{ThM}, we found that, for an arbitrarily small $\e>0$, there exists $n_0$ such
that \baaa \int_{-\pi}^{\pi}|(H_n\ew-1)X_0\ew|d\o\brea
\le \int_{-\pi}^{\pi}|(\wH\ew-H_n\ew)X_0\ew|d\o\\  +
\int_{-\pi}^{\pi}|(\wH\ew-1)X_0\ew|d\o\brea\le
2\pi\e \quad \forall x_0\in\U, \forall n\ge n_0, \eaaa where $\wH=\Z \wh$.

For $\w x_0=\wh\circ x_0$, this implies that
\baaa  \|\w x_0-x_0\|_{\ell_{\infty}}\le\e.\label{eps}\eaaa
\par
Let us estimate the recovery error for the case where $\nuu>0$.
For $\w x=\wh\circ x$, we have that \baaa \|\w x-x\|_{\ell_{\infty}}\le E_0+ E_{\NN},\eaaa
where\baaa &&E_0=\frac{1}{2\pi}\|(H_n\ew-1)X_0\ew|\|_{L_1(-\pi,\pi)}\le \e,\\&&
E_{\NN}=\frac{1}{2\pi}\|(H_n\ew-1)N\ew|\|_{L_1(-\pi,\pi)}.
\eaaa The value $E_{\NN}$ represents the additional error caused by
the presence of unexpected high-frequency noise (when $\nuu>0$). It
follows that \baa \|\w x-x \|_{\ell_{\infty}}\le
\e+\nuu(\kappa+1),\label{yn}\eaa  where
$\kappa=\sup_{\o\in[-\pi,\pi]}|H_n\ew|$.
\par
This means  that the recovery is robust with
respect to noise contamination for any given $\e$.

It can be noted that if $\e\to 0$ then $n\to +\infty$ and $\kappa\to +\infty$. In
this case, error (\ref{yn}) is increasing for any given $\nuu>0$.
This happens when the recovering procedure is targeting too small a size of the
error for the sequences from $\X_\TT$, i.e., under the assumption that
$\nuu=0$.
\noxxx{\par
The equations describing the dependence of $\e$ and $\k$ on $n$
could be derived similarly to estimates in \cite{D12b}, Section 6, obtained  for the predicting problems.}
\section{A numerical example}
\label{SecN}
We made some numerical experiments\xxxonly{ in the spirit of experiments from \cite{D16} but with signals with Z-transform $X\ew$ vanishing at a neigbourhood of  $\o=\pm \pi$  only and }
  with recovering kernels $h_n$ defined in Lemma \ref{lemmaH} for $\TT=\{0,\tau\}$, where $\tau>0$ is an integer.
   As was mentioned above, these kernels  are real valued
even functions on $\ZZ$ such that $\| h_n\|_{\ell_2}\to +\infty$ as $n\to +\infty$.
This means that, for large $n$,  they may decay slow as $|t|\to +\infty$, which would lead to a large error
caused by inevitable data truncation.


 We considered  input  processes $x\in\H$ obtained via the Monte-Carlo simulation
 as the following.
 \begin{enumerate}
 \item At each simulation, a Fourier polynomial $f_1(\o)$ defined on $[-\pi,\pi]$ with $\oo N>0$ non-zero terms with independent
 random coefficients from normal distributions  was created for a given $\oo N>0$.
\item A piecewise continuous function
$f_2(\o)=\sum_{\o\in I_k}\a_k f_1(\o)$ was created for random $\a_k=\xi_k+i\zeta_k$, where $\xi_k$ and $\zeta_k$ were
selected independently from the normal distribution, and where $I_k=(-\pi+(k-1)\pi/\oo N,-\pi+k\pi/\oo N)$, $k=1,...,\oo N$.
\item We defined $X_1\ew=f_2(\o)+\overline{f_2(-\o)}$. This would ensure that  $X_1\ew=\overline{X_1\left(e^{-i\o}\right)}$, $\o\in (-\pi,\pi]$.
\item A input process $x\in\X^\BL$ was obtained using  $X_1$ as
$x=\Z^{-1}\left(\Ind_{\o\in(-\pi,\pi]\setminus \oo D_\e}X_1\ew\right)$ for $\e>0$, where
$\oo D_\e=(-\pi,-\pi+\e)\cup(\pi-\e,\pi)$.
  More precisely, a finite
 set of values for input process $\{x(t)\}_{t=-N}^N$ was calculated for a given $N>0$.
 This represents a truncation of a process  from $\X^\BL\subset\X_\TT$.
\end{enumerate}

We have used R software; the command {\em integrate} was used for calculation
of inverse Z-transforms for $h_n$ and $x$. \index{ It can be noted that, for the particular choice of $X_1$
used in these experiments,  $x$ could be calculated analytically for each Monte-Carlo simulation, i.e. for each   $f_1$ and $\{\a_k\}$. }   We have used $\oo N=10$ and $\e=0.4$.

 Figure \ref{figh}  show example of the traces  of $h_n(t)$ for $\tau=3$ and $n=15$.
Figure \ref{figx}  shows an example of the path for  simulated $x$.

To test our algorithm for   recovery  of missing values $\{x(t)\}_{t=0,\tau}$
from observations of   $\{x(s)\}_{s\neq 0,\tau}$, we calculated their estimates $\w x(t)$  using  convolution with the truncated input $x$
 \baaa
\w x(t)=\sum_{s\in\ZZ:\  |s|\le N}h_n(t-s)x(s).\eaaa
 We calculated the relative error
\baaa
E_{\tau,n}(N,\e)=\EE\frac{\sqrt{\frac{1}{2}\sum_{k=0,\tau}|\w x(k)-x(k)|^2}}{\sqrt{\frac{1}{N}\sum_{t=-N}^N x(t)^2}}.
\eaaa
Here $\EE$ means average over the Monte-Carlo simulations.
In particular, we obtained  that
 \begin{itemize}
 \item[]
$E_{15,15}(150)=0.04$, \quad
$E_{15,15}(300)=0.006$,%
 \item[]
$E_{3,15}(300)=0.2152$, \quad
$E_{3,15}(1500)=0.084$.
\end{itemize}
These examples show that, as expected, the error is decreasing as the truncation parameter  $N$ is increasing if the distance between times $t=\tau$ and $t=0$ for missing values is decreasing. \def\sm{}
\begin{figure}[ht]
\centerline{\epsfig{figure=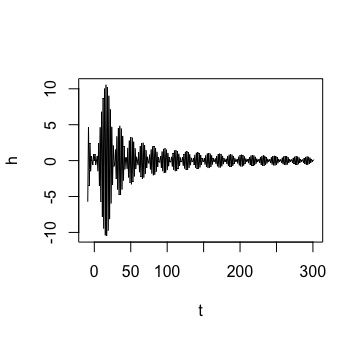,width=9cm,height=6.0cm}}
\vspace{-0.5cm}
\centerline{\epsfig{figure=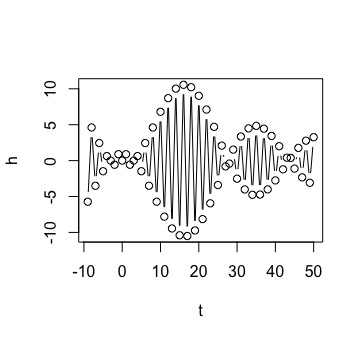,width=9cm,height=6.0cm}}
\caption[]{\sm The kernel $h_n(t)$ for $\TT=\{0,3\}$ and $n=15$, for $t=-10,1,...,300$ and
$t=-10,1,...,50$ (below). }
\label{figh}
\end{figure}
\begin{figure}[ht]
 \centerline{\epsfig{figure=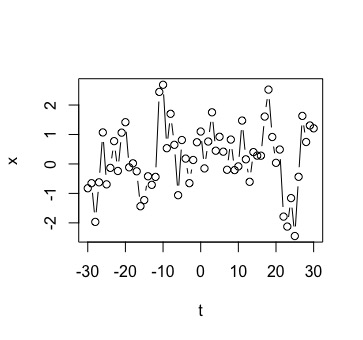,width=9cm,height=6.0cm}}
\caption[]{\sm A trace of $x$ simulated with $\e=0.4$. }
\label{figx}
\end{figure}

\xxxonly{}
\noxxx{}
\end{document}